\documentclass[preprint,prb,aps]{revtex4-2}

\usepackage{graphicx}
\usepackage{dcolumn}
\usepackage{bm} 
\usepackage{natbib}
\usepackage{amsmath,amsfonts}
\usepackage{algorithm}
\usepackage{algpseudocode}
\usepackage{bbm} 
\usepackage{float} 
\usepackage{xcolor}

\DeclareUnicodeCharacter{2062}{\,}

\begin{document}


\title{\boldmath 
Quantum criticality in cuprate superconductors 
revealed by optical conductivity measurement}
\unboldmath

\author{Hwiwoo Park$^{1}$} \author{Sung-Sik Lee$^{2,3}$} \author{G. D. Gu$^{4}$} \author{Jungseek Hwang$^{1,2}$}
\email{jungseek@skku.edu}
\affiliation{$^{1}$Department of Physics, Sungkyunkwan University, Suwon, Gyeonggi-do 16419, Republic of Korea \\  $^{2}$Department of Physics and Astronomy, McMaster University, Hamilton, ON, L8S 4M1, Canada \\ $^{3}$Perimeter Institute for Theoretical Physics, Waterloo, Ontario, N2L 2Y5, Canada \\
$^{4}$Condensed Matter Physics and Materials Science Department, Brookhaven National Laboratory, Upton, NY 11973-5000, USA}

\date{\today}

\begin{abstract}

\noindent {\bf The ubiquitous temperature ($T$)-linear behaviour of the transport scattering rate in the normal state of strongly correlated electron systems is called strange metallicity \cite{zaanen:2004,phillips:2022,hartnoll:2022,chowdhury:2022,yuan:2022}. Although strange metallicity is crucial to understanding superconductivity in correlated electron systems, its origin remains elusive to date \cite{hussey:2023}. Here, we present the doping-, temperature-, and frequency ($\omega$)-dependent transport properties of overdoped Bi$_2$Sr$_2$CaCu$_2$O$_{8+\delta}$ in a wide doping range of 0.183 to 0.231. We observe that the optical scattering rate and effective mass exhibit an $\omega/T$ scaling behaviour at a critical doping of $p_{c} \simeq$ 0.231.
Away from the critical doping, the $\omega/T$ scaling behaviour is destroyed below a doping-dependent crossover temperature $T_\Delta(p) \sim |p-p_{c}|^{0.24}$.
Furthermore, the optical coherence mode (OCM) observed within the superconducting dome rapidly broadens and eventually disappears as the critical doping is approached.
The emergence of the $\omega/T$ scaling behaviour of the transport scattering rate and 
broadening of the OCM near the critical doping strongly suggests that strange metallic behaviour is caused by quantum critical fluctuations. Our results provide compelling spectroscopic evidence for quantum criticality in cuprate superconductors.
}

\end{abstract}

\maketitle


According to the Drude picture, the electrical resistivity of metals is proportional to the rate at which charge carriers experience scatterings that dissipate the current. 
In clean Fermi liquids, the resistivity at low temperatures is controlled by the interaction between quasiparticles, which is suppressed 
as $T^2$ at low temperature ($T$) \cite{uehling:1933,kittel:2005}. 
However, the transport of cuprates in the normal state deviates significantly from the Fermi liquid behaviour. 
Near the optimal doping level yielding the highest superconducting (SC) transition temperature, the resistivity is proportional to $T$.
Although this strange metallic behaviour is ubiquitous in strongly correlated electron systems \cite{zaanen:2004,phillips:2022,hartnoll:2022,chowdhury:2022,yuan:2022}, its origin is not yet fully understood.

The strange metallic behaviour can be explained if one assumes that the scattering rate is only controlled by the Planckian frequency ($k_B T/\hbar$) associated with temperature \cite{zaanen:2004,legros:2019}.
In such Planckian metals, the frequency-dependent scattering rate, which can be measured through the optical conductivity, is expected to be proportional to both $T$ and $\omega$. 
Such a $T$- and $\omega$-linear scattering rate is crucial for the success of the phenomenological model of marginal Fermi liquid (MFL) 
in reproducing various experimental observations \cite{varma:1989,littlewood:1991}.

One possible route for the Planckian dissipation is quantum criticality.
If the correlation length/time of an incipient order parameter is the only scale that controls the scattering rate,
an $\omega/T$ scaling behaviour is expected to emerge within the critical fan that sits above the SC transition temperature over a finite doping range. 
However, the lack of coherent quasiparticles makes it challenging to understand the low-energy physics of such quantum critical metals \cite{lee:2018a}.
Recently, some progress has been made in understanding the strange metallic behaviour in quantum critical metals with certain disorders \cite{patel:2019,patel:2023,li:2024}. 
Experimentally,
a recent optical spectroscopic study \cite{michon:2023} on La$_{2-x}$Sr$_x$CuO$_4$ (LSCO) at the pseudogap critical point ($x = 0.24$) showed that the optical scattering rate and effective mass exhibited $\omega/T$ scaling 
(refer to Methods section for a review of the extended Drude model from which the optical scattering rate and effective mass are defined \cite{gotze:1972,puchkov:1996}). 
The thermodynamic signature of the quantum criticality of LSCO at the pseudogap critical point has been reported in literature \cite{michon:2019}. 
However, the earlier optical study focused on the temperature- and frequency-dependent properties at the single doping level.
Consequently, it has been difficult to verify whether the $\omega/T$ scaling arises due to critical scattering within an extended zero-temperature phase or a quantum critical point.
To resolve this issue, a systematic {\it doping-dependent} optical spectroscopic study of temperature- and frequency-dependent scattering rates is required.

In this study, we examined overdoped Bi$_2$Sr$_2$CaCu$_2$O$_{8+\delta}$ (Bi-2212) single-crystal SC samples in a wide doping range from 0.183 to 0.231 using optical spectroscopy. The complex optical self-energy was obtained from the measured reflectance spectrum using the Kramers-Kronig analysis (refer to Methods) and extended Drude model (refer to Methods). The real part of the optical self-energy exhibits a sharp peak at temperatures near and below the SC transition temperature ($T_c$), which is called the optical coherence mode (OCM). As the doping approaches the highest measured doping level, the coherence mode rapidly widens, and its intensity also rapidly decreases and eventually disappears at the highest doping, which turns out to be the critical doping level, $p_c = 0.234\: (\pm\: 0.004)$. The optical scattering rate and effective mass of the Bi-2212 sample at the highest doping level exhibited $\omega/T$ scaling, as observed in LSCO at the pseudogap critical point \cite{michon:2023}. 
We observed that the scattering rate and effective mass deviate from $\omega/T$ scaling below a doping-dependent crossover temperature ($T_{\Delta}(p)$) that continuously turns on away from the critical doping. 
The best fit of the crossover temperature as a function of doping, that is, $T_{\Delta}(p) \sim |p-p_c|^{z \nu}$, yields $p_c$ = 0.231 ($\pm$ 0.001) and $z \nu = 0.24$ ($\pm\: 0.07$),
where $z\nu$ is the exponent that controls the correlation time ($\xi_\tau$) of the critical fluctuation 
near the critical doping through $\xi_\tau \sim |p-p_c|^{-z \nu}$.
The rapid widening of the OCM and $\omega/T$ scaling near critical doping strongly suggest that the observed critical doping is a quantum critical point (QCP).

\section*{Critical doping: doping-dependent optical self-energy}

We investigated the systematic doping-dependent optical properties of six Bi-2212 single-crystal SC samples in a wide doping range in the overdoped region. The SC transition temperatures ($T_c$) were determined from magnetisation measurements (Extended Data Fig. \ref{figS1}). The determined $T_c$ of the samples were 87, 80, 71, 67, 58, and 53 K in order of increasing doping. The overdoped samples, from the lowest to highest doping, are denoted as OD87, OD80, OD71, OD67, OD58, and OD53. The estimated doping levels of the six Bi-2212 samples using a well-known formula \cite{presland:1991} were 0.183, 0.198, 0.212, 0.217, 0.226, and 0.231, from the lowest to highest doped samples, respectively. The measured reflectance and optical conductivity spectra of the six Bi-2212 samples are shown in Extended Data Fig. \ref{figS2}. A brief description of the $T$- and doping-dependent properties of these two optical quantities is provided in the Methods section. 
In the extended Drude model (refer to Methods), 
the complex optical self-energy is defined as 
$
-2\tilde{\Sigma}^{op}(\omega, T ) \equiv -2\Sigma^{op}_1(\omega, T)-
2i \Sigma^{op}_2(\omega, T) 
 = i(\Omega_p^2/4\pi)\: [1/\tilde{\sigma}(\omega, T)]-\omega$,
where 
$\Sigma^{op}_1(\omega, T)$ and $\Sigma^{op}_2(\omega, T)$ denote the real and imaginary parts of the self-energy, $\tilde{\sigma}(\omega, T)$ is the complex optical conductivity,
and
$\Omega_p$ is the plasma frequency contributed from itinerant charge carriers. 
The real part of the optical self-energy  is closely related to the optical effective mass ($m_{op}^*(\omega, T)$) through
 $-2\Sigma_1^{op}(\omega, T) = \omega[m_{op}^*(\omega, T)/m_b-1]$, where $m_b$ is a band (or undressed) mass.
On the other hand, the imaginary part of the optical self-energy is related to the optical scattering rate ($1/\tau^{op}(\omega, T)$), that is, $-2\Sigma_2^{op}(\omega, T) = 1/\tau^{op}(\omega, T)$.

\begin{figure}[!htbp]
\begin{center}
  \vspace*{-0.3 cm}%
\includegraphics[width=1.0 \columnwidth]{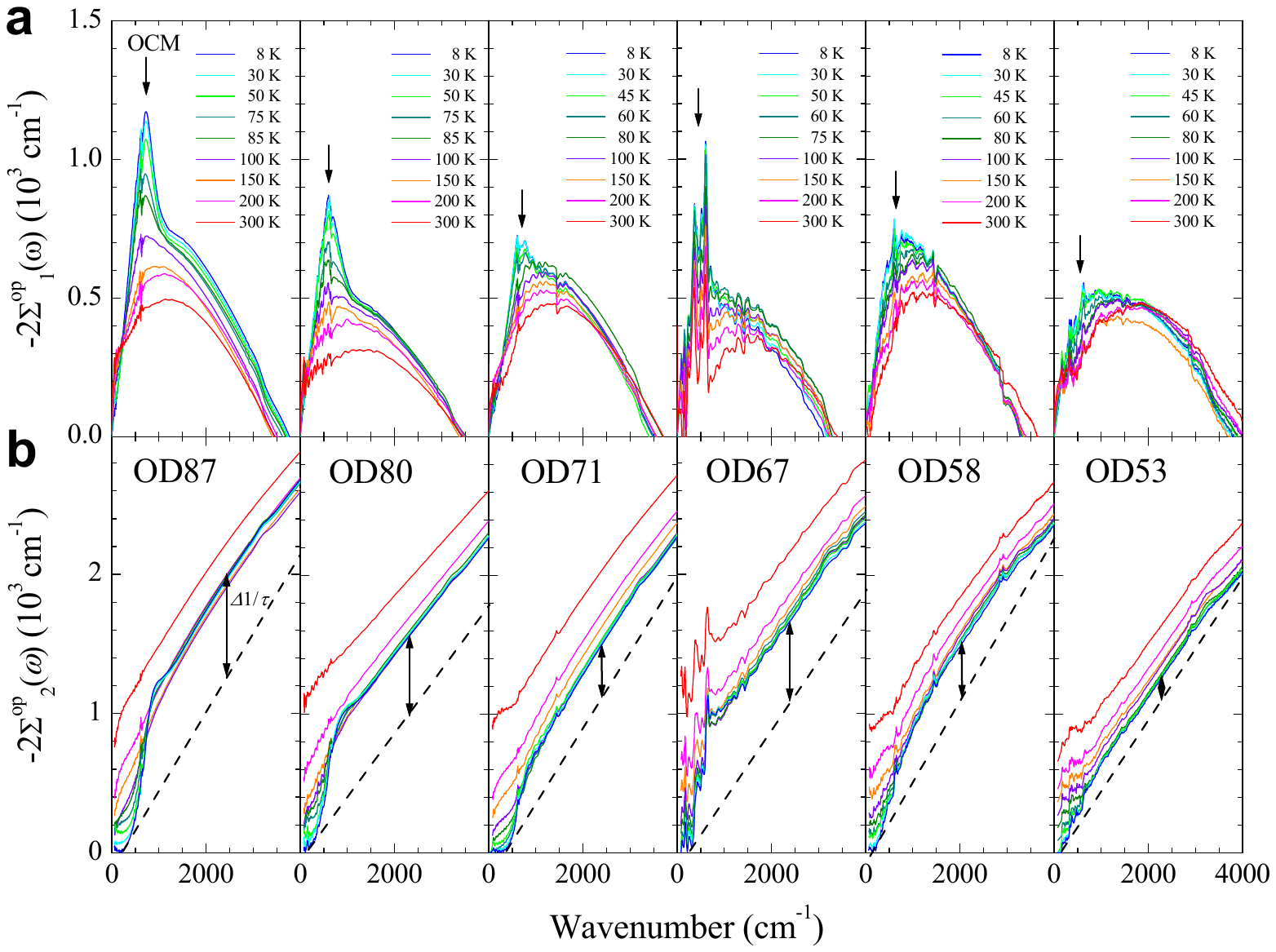}
  \vspace*{-1.5 cm}%
\caption{{\bf Doping-dependent evolution of optical self-energy.} (a) Real parts of the six Bi-2212 samples at various temperatures above and below $T_c$. The vertical arrows indicate the OCM. (b) Corresponding imaginary parts (or optical scattering rate ($1/\tau^{op}(\omega)$) of the six Bi-2212 samples. $\Delta 1/\tau$ indicates the rise of the step-like feature. The black dashed lines are used for estimating the $\Delta 1/\tau$. 
\label{fig1}}
\end{center}
\end{figure}

The real parts of the optical self-energies of all six Bi-2212 samples at various temperatures above and below $T_c$ are shown in Fig. \ref{fig1}a. At 300 K, the real part of the optical self-energy exhibits a single broad peak. As the temperature decreases, a sharp peak appears and grows (marked with an arrow) near and below $T_c$. 
This sharp peak weakens as the doping increases and is known as the OCM \cite{hwang:2004,hwang:2007a}. The OCM is closely related to a sharp peak in the nodal direction spectrum observed in a previous angle-resolved photoemission spectroscopy (ARPES) study \cite{johnson:2001} at low temperatures near and below $T_c$. Moreover, the OCM is related to magnetic resonance modes \cite{carbotte:1999,hwang:2006,hwang:2008c}. The real parts of the optical self-energies of the six overdoped Bi-2212 samples and an optimally doped Bi-2212 from a previous study \cite{hwang:2007a} are shown in the top panels of Extended Data Fig. \ref{figS3}. 
The corresponding optical effective masses ($m_{op}^*(\omega, T)/m_b$) are shown in the middle panels of Extended Data Fig. \ref{figS3}. 

\begin{figure}[!htbp]
\begin{center}
  \vspace*{-1.0 cm}%
\includegraphics[width=1.0 \columnwidth]{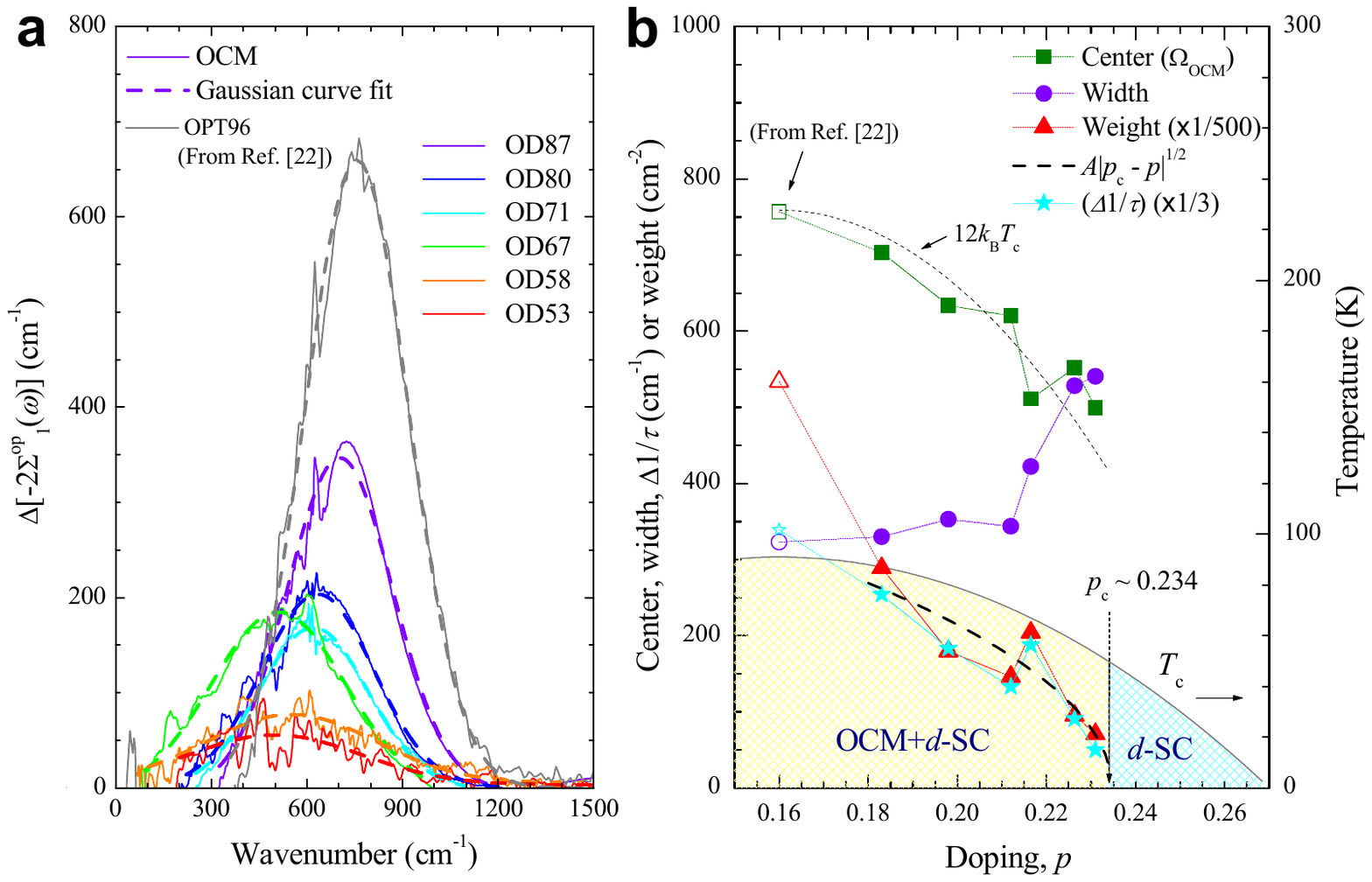}
  \vspace*{-2.4 cm}%
\caption{{\bf OCM and critical doping.} (a) OCM separated from the high-temperature background of the six Bi-2212 samples and one optimally doped Bi-2212 from a previous study \cite{hwang:2007a} and their Gaussian curve fits. (b) Doping-dependent fitting parameters of the OCM. The doping-dependent rise ($\Delta 1/\tau$) of the step-like feature in the optical scattering rate is also displayed (see Fig. \ref{fig1}b).
\label{fig2}}
\end{center}
\end{figure}

To examine the doping dependence of the OCM, the coherence mode at the lowest $T$ below $T_c$ was separated by subtracting the high-temperature background, that is, $\Delta [-2\Sigma_1^{op}(\omega)] \equiv [-2\Sigma_1^{op}(\omega, 8\: \mbox{K})] -[-2\Sigma_1^{op}(\omega, T \simeq T_c)] $. In Fig. \ref{fig2}a, we show the separated coherence modes and their Gaussian profile fits at various doping levels. Here, we include the spectrum of the optimally doped Bi-2212 sample from a previous optical study \cite{hwang:2007a}. 
The mode frequency, width, and weight of the peak as functions of doping are shown in Fig. \ref{fig2}b.
The centre frequency ($\Omega_{OCM}$) is approximately proportional to $T_c$, that is, $\Omega_{OCM} \sim 12 k_B T_c$ (thin black dashed line). 
However, the width and weight of the peak exhibit more drastic doping dependences.
As doping increases, the width of OCM increases slightly up to $p \sim$ 0.212 and then rapidly increases above it. 
The OCM also weakens rapidly above $p$ $\sim$ 0.212 and eventually disappears at a critical doping ($p_{c}$) inside the SC dome. We estimated the critical doping by fitting the weight with $A_0(p_{c}-p)^{1/2}$ (thick black dashed line). The estimated fitting parameters are the amplitude, $A_0 =$ 1152 $(\pm\: 140)$ cm$^{-2}$, and the critical doping, $p_{c}$ = 0.234 ($\pm$ 0.004). The critical doping divides the SC dome into two parts with and without the coherence mode. 
Note that an earlier optical study on Tl$_2$Ba$_2$CuO$_{6+\delta}$ (Tl-2201) observed that the OCM completely disappeared above the critical doping ($p_{c}$) \cite{ma:2006}, although it is a different cuprate system.

The doping-dependent evolution of OCM is also reflected in the imaginary part of the optical self-energy.
The optical scattering rates of the six Bi-2212 samples at various temperatures are shown in Fig. \ref{fig1}b. 
As the temperature decreases near and below $T_c$, a step-like feature directly associated with the OCM develops at approximately 1000 cm$^{-1}$. We denote the maximum size of the step-like feature as $\Delta 1/\tau$, as shown in the figure. As the doping increases, $\Delta 1/\tau$ decreases, and its doping dependence perfectly tracks that of the weight of OCM, as shown in Fig. \ref{fig2}b. 
Note that the OD67 sample shows strong temperature-independent sharp peaks between 300 and 700 cm$^{-1}$, the origin of which is not known yet; the coherence mode and step-like feature were influenced by the sharp peaks. 
Extended Data Fig. \ref{figS4} shows a $T$-dependent evolution of the optical self-energy over a wide range of doping, where its doping-dependent evolution at given temperatures is clearly observed.

While the doping-dependent spectral weight of OCM has been observed within a limited range of doping 
in previous studies \cite{hwang:2004,hwang:2007a},
our systematic study of highly overdoped samples shows that OCM rapidly widens approaching the critical doping (refer to Extended Data Fig. \ref{figS5}) and disappears at the critical doping. 
This suggests the presence of strong quantum fluctuations that make OCM incoherent near the critical doping.
In the rest of the paper, we provide evidence for the existence of a QCP by examining the optical scattering rate and effective mass in the normal state.

\section*{Quantum criticality: $\omega/T$ scaling in the optical scattering rate and effective mass}

\begin{figure}[!htbp]
\begin{center}
  \vspace*{-0.3 cm}%
\includegraphics[width=1.0 \columnwidth]{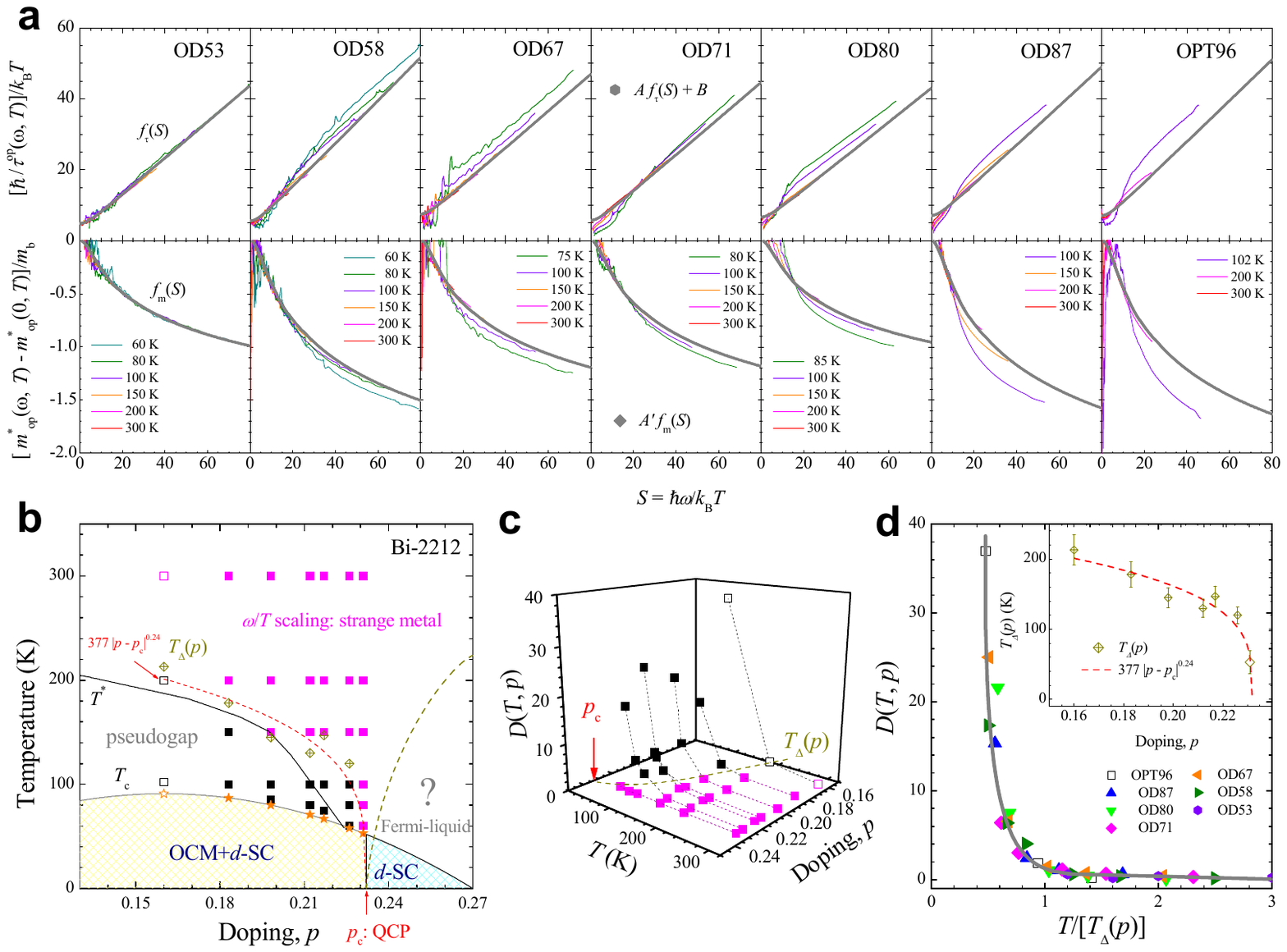}
  \vspace*{-1.3 cm}%
\caption{{\bf $\omega/T$ scaling and quantum criticality.} (a) (Upper panels) $[\hbar/\tau^{op}(\omega, T)]/k_B T$ versus $S$ curves of all the six Bi-2212 samples and an optimally doped Bi-2212 from a previous study \cite{hwang:2007a} with $\omega <$ 0.4 eV. The gray solid hexagon symbol is $A f_{\tau}(S)+B$ for each sample, where $f_{\tau}(S)$ is the scaling function at $p$ = 0.231. (Lower panels) $[m^*_{op}(\omega, T)-m^*_{op}(0, T)]/m_b$ versus $S$ of all six samples and an optimally doped Bi-2212 from a previous study \cite{hwang:2007a} with $\omega <$ 0.4 eV. The gray solid diamond symbol is $A' f_{m}(S)$ for each sample, where $f_{m}(S)$ is the scaling function at $p$ = 0.231. Note that all the spectra are in the normal state. (b) Phase diagram of Bi-2212 in the overdoped region. The $T^*(p)$ line is brought from a previous Raman study \cite{auvray:2019} and $T_{\Delta}(p)$ is the deviation temperature from $\omega/T$ scaling as a function of doping. (c) Normalized deviations from the fits, $D(T, p)$, for all seven samples, including the optimally doped one, are displayed. $T_{\Delta}(p)$ of each sample is estimated based on $D(T, p)$. (d) $D(T,p)$ versus $T/[T_{\Delta}(p)]$ curves of all seven samples collapse into a single curve. The thick line is a guide to eyes. In the inset, the $T_{\Delta}(p)$ and a fit with $377 \:|p-p_c|^{0.24}$ and $p_c$ = 0.231 are shown.
\label{fig3}}
\end{center}
\end{figure}

We first consider the critical doping ($p_c=0.231$).
To check the critical scaling of the optical scattering rate and the effective mass,
we plot $[\hbar/\tau^{op}(\omega, T)]/k_B T$
and
$[m^*_{op}(\omega, T)-m^*_{op}(0, T)]/m_b$ 
for data measured at various temperatures above $T_c$
as functions of $S \equiv \hbar \omega/k_BT $ 
in the upper 
and the lower panels of Fig. \ref{fig3}a, respectively (see the left most panels). 
We use the same energy cutoff (0.4 eV) used in previous literature to include only the contribution of the charge carriers \cite{michon:2023}.
Note that it is not simple to determine $m^*_{op}(0, T)/m_b$ due to large uncertainty at low frequencies. The determined $m^*_{op}(0, T)/m_b$ is shown in Extended Data Fig. \ref{figS6}. At the critical doping, $m^*_{op}(0, T)]/m_b$ is proportional to $-\log(T)$.
Both $[\hbar/\tau^{op}(\omega, T)]/k_B T$
and
$[m^*_{op}(\omega, T)-m^*_{op}(0, T)]/m_b$ 
exhibit excellent data collapses.
This $\omega/T$ scaling behaviour is in agreement with the previous optical study done on LSCO at the pseudogap critical doping level \cite{michon:2023}. 
Furthermore, the $T$-linear scattering rate and
the effective mass that scales as $-\log(T)$ are consistent with the MFL behaviour\cite{varma:1989,littlewood:1991}. 

In order to tell whether the scaling behaviour is due to a quantum critical point or an extended critical phase, we now examine how $\omega/T$ scaling changes with doping.
In Fig. \ref{fig3}a, 
$[\hbar/\tau^{op}(\omega, T)]/k_B T$ as a function of $S$ are fitted with $A f_{\tau}(S) + B$
for all seven Bi-2212 samples. 
Here, $f_{\tau}(S)$ is the scaling function 
that fits the collapsed data at the critical doping. 
$A$ and $B$ are doping-dependent parameters adjusted to fit the data at 300 K 
that exhibit $\omega/T$ scaling even away from the critical doping.  
The fits are shown in the upper panels of Fig. \ref{fig3}a and the fitting parameters ($A$ and $B$) are shown in Extended Data Fig. \ref{figS7}. 
Away from the critical doping, the $T$-linear scattering rate is replaced with the $T^2$ behaviour of Fermi liquids below a crossover temperature $T_\Delta(p)$. A previous optical study on cuprates in the pseudogap phase presented spectroscopic evidence for the Fermi liquid behaviour \cite{mirzaei:2013}. 
The violation of the  $\omega/T$ scaling at low temperatures
can be quantified through
$D(T, p) = [1/S_c] \int_0^{S_c} \{ [\hbar/\tau^{op}(T, p, S)]/k_B T - [A(p)f_{\tau}(S) + B(p)] \}^2 dS$
that measures the normalized deviation from the scaling function,
where $S_c$ is a cutoff.
We identify $T_{\Delta}(p)$ as the temperature below which $D(T, p)$ rapidly grows, as shown in Fig. \ref{fig3}c. 
$T_{\Delta}(p)$ divides the strange metallic state with $\omega/T$ scaling (magenta squares) from Fermi liquids (black squares).
The
dimensionless $D(T, p)$ values depend only on the reduced temperature $T/[T_{\Delta}(p)]$, collapsing into a single $T/[T_{\Delta}(p)]$ scaling curve, as shown in Fig. \ref{fig3}d.
$T_{\Delta}(p)$ can be interpreted as the energy scale of a mode that becomes critical at the critical doping.
$T_{\Delta}(p)$ is well fitted by $T_0\:|p-p_c|^{z \nu}$ with $T_0$ = 377 ($\pm$ 91) K, the critical doping, $p_c$ = 0.231 ($\pm$ 0.001), and a critical exponent, $z \nu$ = 0.24 ($\pm$ 0.07), as shown in the inset of Fig. \ref{fig3}d. Note that the critical doping levels estimated from the weight of OCM (Fig. \ref{fig2}b) and $T_{\Delta}(p)$ are basically identical within the error range.
An earlier theoretical study reported that the combined critical exponent, $z\nu$, of the strange metallic region of cuprate is 0.67 based on five different DC transport quantities \cite{hartnoll:2015}. 
In our optical spectroscopic study, we obtained the combined exponent, 0.24, which is significantly smaller than that reported previously.

Remarkably, $T_{\Delta}(p)$ is close to the pseudogap temperature ($T^*$) reported previously \cite{auvray:2019}, as shown in Fig. \ref{fig3}b. 
$T_{\Delta}(p)$ is also close to the onset temperature of the OCM
because the step-like feature associated with OCM causes deviation from the $\omega/T$ scaling.
This suggests that the quantum critical fluctuations associated with the melting of the order that causes the pseudogap behaviour is responsible for the onset of the strange metallic behaviour and the destruction of OCM within a critical fan above $T_{\Delta}(p)$.

Similarly, in the lower panels of Fig. \ref{fig3}a, $[m^*_{op}(\omega, T)-m^*_{op}(0, T)]/m_b$ versus $S$ curves of all seven Bi-2212 samples were also fitted with $A'f_m(S)$ with the same assumption as for the scattering rate, where $f_m(S)$ is the scaling function for the effective mass
that fits the collapsed data at the critical doping.
The fitting parameter $A'$ is shown in Extended Data Fig. \ref{figS7}. 
Note that the $[m^*_{op}(\omega, T)-m^*_{op}(0, T)]/m_b$ versus $S$ curves are less reliable than the $[\hbar/\tau^{op}(\omega, T)]/k_B T$ versus $S$ curves due to the uncertainties in $m^*_{op}(0, T)]/m_b$.

\section*{Discussion and outlook}

Our results provide compelling spectroscopic evidence for quantum criticality in overdoped Bi-2212. 
However, the origin of the OCM that enables us to probe the criticality is still unclear.
One possible origin of the OCM is the Leggett mode, which is a theoretically proposed collective mode that can be observed in multiband superconductors by optical absorption \cite{kamatani:2022}. In the presence of the pair density wave (PDW), observed in the pseudogap and superconducting phases of an underdoped Bi-2212 \cite{du:2020,wang:2021},
the Leggett mode can be optically activated in the linear response regime \cite{nagashima:2025}.
Even in the absence of the PDW order, cuprates may support what can be broadly viewed as Leggett modes that describe the relative phase fluctuations between two orthogonal anti-nodal directions.

If the OCM mode is the $d$-wave Leggett mode, one may be able to explain its broadening near the critical doping rather naturally.
Previous experimental studies \cite{auvray:2019,kaminski:2006} revealed two critical doping levels of Bi-2212 in the overdoped region, similar to the critical doping level in this study. 
The first represents nematic fluctuations maximized near the end point of the pseudogap phase at $p^* \cong$ 0.22 observed by Raman spectroscopy \cite{auvray:2019}. 
The second is the Fermi-surface topology change of the antibonding band from hole-like to electron-like, exhibiting a Lifshitz transition at a critical doping ($p \cong$ 0.225) observed by ARPES \cite{kaminski:2006}. 
However, it may well be that there is only one critical doping associated with the Lifshitz transition,
which enhances nematic fluctuations \cite{loret:2018,gupta:2021}. 
Given that the nematic mode and the $d$-wave Leggett mode have the same symmetry in the $d$-wave superconductor, the latter is expected to become incoherent by enhanced nematic fluctuations.
To test this idea, it will be interesting to see how the uniaxial pressure, which suppresses nematic fluctuations, affects the broadening of OCM.
It will also be interesting to understand the relation between the OCM and the magnetic resonance mode, which is observed through various spectroscopic techniques in cuprates, including Bi-2212 \cite{carbotte:1999,johnson:2001,hwang:2004,hwang:2006,carbotte:2011,bourges:2005,yu:2009}. Identifying the microscopic relationship between them will aid in determining the origin of $\omega/T$ scaling and the quantum criticality, eventually elucidating the mechanism of high-$T_c$ superconductivity.

The last point we would like to address is that the doping coverage of the bulk Bi-2212 is limited by the overdoped region inside the SC dome. Therefore, further optical spectroscopic studies using other cuprate systems may be necessary to cover wider and higher doping regions. Candidate cuprate systems are single-layer Bi-based cuprate (Bi$_2$Sr$_{2-x}$La$_x$CuO$_{6+\delta}$), LSCO, Tl-2201, and Nd-doped LSCO. Systematic temperature- and doping-dependent optical spectroscopic studies on these candidate cuprates will provide further information on $\omega/T$ scaling (or quantum fluctuations), quantum criticality, extended quantum criticality \cite{cooper:2009}, and highly overdoped Fermi liquid behaviour.

\newpage

\section*{Methods}

\subsection*{Single-crystal Bi-2212 samples and $T$-dependent reflectance measurement}

Single-crystal Bi-2212 samples were grown using the travelling solvent floating zone technique at the Brookhaven National Laboratory. The SC transition temperatures ($T_c$) of the samples were determined using magnetisation measurements (Extended Data Fig. \ref{figS1}). The determined $T_c$ of the samples were 87, 80, 71, 67, 58, and 53 K in the order of increasing doping. The samples from the lowest to the highest doping levels are denoted as OD87, OD80, OD71, OD67, OD58, and OD53. We estimated the doping levels of these Bi-2212 samples using a well-known formula \cite{presland:1991} as $p(T_c) = 0.16 \mp [(1/82.6)(1- T_c/T_c^{max}) ]^{1/2}$, where $T_c^{max}$ is the maximum $T_c$ at the optimal doping and $\mp$ denotes underdoped and overdoped cases, respectively. The estimated doping levels of the six Bi-2212 samples were 0.183, 0.198, 0.212, 0.217, 0.226, and 0.231, from the lowest to highest doping levels, respectively. We measured the reflectance spectra of the samples in a wide spectral range from the far-infrared to ultraviolet regions (80-40000 cm$^{-1}$) at various temperatures from 8 K to 300 K using a commercial Fourier-transform infrared spectrometer (Vertex 80v, Bruker, Germany) and a temperature-dependent measurement system (LT3B Helitran Advanced Research Systems, USA). We used an unpolarised beam for the $ab$-plane reflectance measurement, with an incident angle on the sample of $\sim$10$^{\:\mathrm{o}}$. To obtain an accurate reflectance spectrum, we utilised an {\it in situ} gold/aluminium evaporation method \cite{homes:1993}. In this method, we used a 200 nm-thick gold or aluminium film coated on the sample as a reflectance reference. Gold was used in the far- and mid-infrared regions, whereas aluminium was used in the near-infrared, visible, and ultraviolet regions. Furthermore, we corrected the measured reflectance with respect to the gold or aluminium film by multiplying in with the absolute reflection of the gold or aluminium film. 

\subsection*{Kramers-Kronig analysis}

Kramers-Kronig analysis \cite{tanner:2019} was employed on the measured reflectance spectra to obtain optical constants, such as the optical conductivity. To perform the Kramers-Kronig analysis, the measured reflectance spectrum ($R(\omega)$) in a finite spectral range must be extrapolated to both sides: zero and infinity. For extrapolation from the lowest data point to zero frequency, the Hagen-Rubens relation, that is, $1-R(\omega) \propto \sqrt{\omega}$, was used for the normal state, and $1-R(\omega) \propto \omega^4$ for the SC state. For extrapolation from the highest data point to infinity for both states, up to 10$^6$ cm$^{-1}$, $R(\omega) \propto \omega^{-1}$ was used, and above 10$^6$ cm$^{-1}$, the free-electron behaviour, that is, $R(\omega) \propto \omega^{-4}$, was assumed.

\subsection*{Reflectance and optical conductivity}

The measured $ab$-plane reflectance spectra of the six samples at various temperatures are shown in the top and middle panels of Extended Data Fig. \ref{figS2}. The overall reflectance increased with doping. All the samples exhibited metallic behaviour; at low frequencies, the reflectance increased as the temperature decreased. Near or below $T_c$, a shoulder-like feature appeared near 1000 cm$^{-1}$ and became weaker with increasing doping. The complex optical conductivity spectrum ($\tilde{\sigma}(\omega) \equiv \sigma_1(\omega)+i\sigma_2(\omega)$) corresponding to the measured reflectance spectrum was obtained using the Kramers-Kronig analysis \cite{tanner:2019}. The real parts of the optical conductivities of all six Bi-2212 samples at various temperatures are shown in the bottom panels of Extended Data Fig. \ref{figS2}. The optical conductivity in the normal state shows a Drude-like peak near the zero frequency. Near or below $T_c$, a dip feature appears below 1000 cm$^{-1}$. The dip feature is associated with the SC gap and the sharp peak in the bosonic spectrum \cite{hwang:2007} and becomes weaker as the doping increases. To study the dip features and correlations between the charge carriers in more detail, the measured spectra were further analysed using the extended Drude model formalism (refer to Methods).

\subsection*{Extended Drude model and optical self-energy}

The extended Drude model is a generalisation of the simple Drude model achieved by replacing the $\omega$-independent scattering rate with the $\omega$-dependent one and, therefore, is used to describe the correlated electrons. The complex optical conductivity ($\tilde{\sigma}(\omega, T) \equiv \sigma_1(\omega, T)+i\sigma_2(\omega, T)$) in the extended Drude model is written as follows \cite{gotze:1972,puchkov:1996,hwang:2004}:
\begin{equation}
  \tilde{\sigma}(\omega, T) = i\frac{\Omega_p^2}{4 \pi} \frac{1}{\omega + [-2\tilde{\Sigma}^{op}(\omega, T)]},
\end{equation}
where $\Omega_p$ is the plasma frequency and $\tilde{\Sigma}^{op}(\omega, T) $ $(\equiv \Sigma^{op}_1(\omega, T) +i\Sigma^{op}_2(\omega, T) $) is the complex optical self-energy that carries correlation information. Moreover, $\Omega_p^2 (= 4 \pi n e^2/m_b) $ is proportional to the charge carrier density ($n$), where $e$ is the unit charge and $m_b$ is the band (or undressed) mass. The optical self-energy is a two-particle self-energy because electrons (occupied state) and holes (unoccupied state) are involved in the optical absorption process. The real part ($\Sigma^{op}_1(\omega, T)$) is related to the optical coupling strength ($\lambda^{op}(\omega, T)$) as $-2\Sigma^{op}_1(\omega, T) = \omega \lambda^{op}(\omega, T) = (\Omega_p^2/4\pi)\: \sigma_2(\omega,T)/\{[\sigma_1(\omega,T)] ^2+ [\sigma_2(\omega, T)] ^2\}-\omega$, and also to the optical effective mass ($m^{*}_{op}(\omega, T)$) as $-2\Sigma^{op}_1(\omega, T) = \omega [m^{*}_{op}(\omega, T)/m_b - 1] $. The imaginary part ($\Sigma^{op}_2(\omega, T)$) is related to the optical scattering rate ($1/\tau^{op}(\omega, T)$), that is, $-2\Sigma^{op}_2(\omega, T) = 1/\tau^{op}(\omega, T) = (\Omega_p^2/4 \pi)\: \sigma_1(\omega,T)/\{[\sigma_1(\omega,T)]^2+ [\sigma_2(\omega, T)]^2\}$. Note that the real and imaginary parts form a Kramers-Kronig pair. To obtain the optical self-energy from the optical conductivity, the plasma frequency ($\Omega_p$) contributed only by the intraband transitions (or charge carriers) is required and can be estimated by using a previously proposed method \cite{hwang:2007a}. The estimated plasma frequencies ($\Omega_p$) of the OD87, OD80, OD71, OD67, OD58, and OD53 samples are 16018, 18290, 18717, 18343, 18471, and 17918 cm$^{-1}$, respectively. One also needs to include the background dielectric constant ($\varepsilon_H$), contributed from the high-frequency absorptions above the intraband transitions, which can be estimated using a previously proposed method \cite{hwang:2007a}. The estimated background dielectric constants ($\varepsilon_H$) of the OD87, OD80, OD71, OD67, OD58, and OD53 samples are 4.25, 4.83, 4.94, 5.39, 5.77, and 4.04, respectively.

\acknowledgments  We thank Tuson Park and Soon-Gil Jung for helping us measure the magnetic susceptibility. This study was supported by the National Research Foundation of Korea (NRFK Grants numbers 2021R1A2C101109811 and RS-2024-00460248). Research at the Perimeter Institute was supported in part by the Government of Canada through Industry Canada and by the Province of Ontario through the Ministry of Research and Information. S.-S.L. acknowledges support from the Natural Sciences and Engineering Research Council of Canada.\\ 

{\bf AUTHOR CONTRIBUTION}\\
H.P. and J.H. measured and analysed the optical spectra. J.H. wrote the draft. S.-S.L. and J.H. discussed the measured spectra for further analysis, built the physical picture, and edited the manuscript. G.D.G. provided Bi-2212 single-crystal samples. All the authors contributed to the preparation of this paper.\\

{\bf COMPETING INTERESTS} \\
The authors declare no competing interests.\\

{\bf DATA AVAILABILITY}\\
Source data will be provided with reasonable requests.\\

{\bf Correspondence} and requests for materials should be addressed to jungseek@skku.edu.

\bibliographystyle{naturemag}
\bibliography{bib.bib}

\begin{thebibliography}{10}
\expandafter\ifx\csname url\endcsname\relax
  \def\url#1{\texttt{#1}}\fi
\expandafter\ifx\csname urlprefix\endcsname\relax\def\urlprefix{URL }\fi
\providecommand{\bibinfo}[2]{#2}
\providecommand{\eprint}[2][]{\url{#2}}

\bibitem{zaanen:2004}
\bibinfo{author}{Zaanen, J.}
\newblock \bibinfo{title}{Why the temperature is high}.
\newblock \emph{\bibinfo{journal}{Nature}} \textbf{\bibinfo{volume}{430}},
  \bibinfo{pages}{512} (\bibinfo{year}{2004}).

\bibitem{phillips:2022}
\bibinfo{author}{Phillips, P.~W.}, \bibinfo{author}{Hussey, N.~E.} \&
  \bibinfo{author}{Abbamonte, P.}
\newblock \bibinfo{title}{Stranger than metals}.
\newblock \emph{\bibinfo{journal}{Science}} \textbf{\bibinfo{volume}{377}},
  \bibinfo{pages}{eabh4273} (\bibinfo{year}{2022}).

\bibitem{hartnoll:2022}
\bibinfo{author}{Hartnoll, S.~A.} \& \bibinfo{author}{Mackenzie, A.~P.}
\newblock \bibinfo{title}{Colloquium: Planckian dissipation in metals}.
\newblock \emph{\bibinfo{journal}{Rev. Mod. Phys.}}
  \textbf{\bibinfo{volume}{94}}, \bibinfo{pages}{041002}
  (\bibinfo{year}{2022}).

\bibitem{chowdhury:2022}
\bibinfo{author}{Chowdhury, D.}, \bibinfo{author}{Georges, A.},
  \bibinfo{author}{Parcollet, O.} \& \bibinfo{author}{Sachdev, S.}
\newblock \bibinfo{title}{\mbox{Sachdev-Ye-Kitaev} models and beyond: Window
  into non-fermi liquids}.
\newblock \emph{\bibinfo{journal}{Rev. Mod. Phys.}}
  \textbf{\bibinfo{volume}{94}}, \bibinfo{pages}{035004}
  (\bibinfo{year}{2022}).

\bibitem{yuan:2022}
\bibinfo{author}{Yuan, J.} \emph{et~al.}
\newblock \bibinfo{title}{Scaling of the strange-metal scattering in
  unconventional superconductors}.
\newblock \emph{\bibinfo{journal}{Science}} \textbf{\bibinfo{volume}{602}},
  \bibinfo{pages}{431} (\bibinfo{year}{2022}).

\bibitem{hussey:2023}
\bibinfo{author}{Hussey, N.~E.}
\newblock \bibinfo{title}{High-temperature superconductivity and strange
  metallicity: Simple observations with (possibly) profound implications}.
\newblock \emph{\bibinfo{journal}{Physca C: Superconductivity and its
  applications}} \textbf{\bibinfo{volume}{614}}, \bibinfo{pages}{1354362}
  (\bibinfo{year}{2023}).

\bibitem{uehling:1933}
\bibinfo{author}{Uehling, E.~A.} \& \bibinfo{author}{Uhlenbeck, G.~E.}
\newblock \bibinfo{title}{Transport phenomena in einstein-bose and fermi-dirac
  gases. i}.
\newblock \emph{\bibinfo{journal}{Phys. Rev.}} \textbf{\bibinfo{volume}{43}},
  \bibinfo{pages}{552} (\bibinfo{year}{1933}).

\bibitem{kittel:2005}
\bibinfo{author}{Kittel, C.}
\newblock \emph{\bibinfo{title}{Introduction to solid state physics}}
  (\bibinfo{publisher}{Willey, New York}, \bibinfo{year}{2005}).

\bibitem{legros:2019}
\bibinfo{author}{Legros, A.} \emph{et~al.}
\newblock \bibinfo{title}{Universal \mbox{T}-linear resistivity and planckian
  dissipation in overdoped cuprates}.
\newblock \emph{\bibinfo{journal}{Nat. Phys.}} \textbf{\bibinfo{volume}{15}},
  \bibinfo{pages}{142} (\bibinfo{year}{2019}).

\bibitem{varma:1989}
\bibinfo{author}{Varma, C.}, \bibinfo{author}{Littlewood, P.},
  \bibinfo{author}{Schmitt-Rink, S.}, \bibinfo{author}{Abrahams, E.} \&
  \bibinfo{author}{Ruckenstein, A.}
\newblock \bibinfo{title}{Phenomenology of the normal state of \mbox{Cu-O}
  high-temperature superconductors}.
\newblock \emph{\bibinfo{journal}{Phys. Rev. Lett.}}
  \textbf{\bibinfo{volume}{63}}, \bibinfo{pages}{1996} (\bibinfo{year}{1989}).

\bibitem{littlewood:1991}
\bibinfo{author}{Littlewood, P.~B.} \& \bibinfo{author}{Varma, C.~M.}
\newblock \bibinfo{title}{Phenomenology of the normal and superconducting
  states of a marginal fermi liquid (invited)}.
\newblock \emph{\bibinfo{journal}{J. Appl. Phys.}}
  \textbf{\bibinfo{volume}{69}}, \bibinfo{pages}{4979} (\bibinfo{year}{1991}).

\bibitem{lee:2018a}
\bibinfo{author}{Lee, S.-S.}
\newblock \bibinfo{title}{Recent developments in non-fermi liquid theory}.
\newblock \emph{\bibinfo{journal}{Annual Review of Condensed Matter Physics}}
  \textbf{\bibinfo{volume}{9}}, \bibinfo{pages}{227--244}
  (\bibinfo{year}{2018}).

\bibitem{patel:2019}
\bibinfo{author}{Patel, A.~A.} \& \bibinfo{author}{Sachdev, S.}
\newblock \bibinfo{title}{Theory of a planckian metal}.
\newblock \emph{\bibinfo{journal}{Phys. Rev. Lett.}}
  \textbf{\bibinfo{volume}{123}}, \bibinfo{pages}{066601}
  (\bibinfo{year}{2019}).

\bibitem{patel:2023}
\bibinfo{author}{Patel, A.~A.}, \bibinfo{author}{Guo, H.},
  \bibinfo{author}{Esterlis, I.} \& \bibinfo{author}{Sachdev, S.}
\newblock \bibinfo{title}{Universal theory of strange metals from spatially
  random interactions}.
\newblock \emph{\bibinfo{journal}{Science}} \textbf{\bibinfo{volume}{381}},
  \bibinfo{pages}{790} (\bibinfo{year}{2023}).

\bibitem{li:2024}
\bibinfo{author}{Li, C.} \emph{et~al.}
\newblock \bibinfo{title}{Strange metal and superconductor in the
  two-dimensional yukawa-sachdev-ye-kitaev model}.
\newblock \emph{\bibinfo{journal}{Phys. Rev. Lett.}}
  \textbf{\bibinfo{volume}{133}}, \bibinfo{pages}{186502}
  (\bibinfo{year}{2024}).

\bibitem{michon:2023}
\bibinfo{author}{Michon, B.} \emph{et~al.}
\newblock \bibinfo{title}{Reconciling scaling of the optical conductivity of
  cuprate superconductors with planckian resistivity and specific heat}.
\newblock \emph{\bibinfo{journal}{Nat. Comm.}} \textbf{\bibinfo{volume}{14}},
  \bibinfo{pages}{3033} (\bibinfo{year}{2023}).

\bibitem{gotze:1972}
\bibinfo{author}{G\"{o}tze, W.} \& \bibinfo{author}{W\"{o}lfle, P.}
\newblock \bibinfo{title}{Homogeneous dynamical conductivity of simple metals}.
\newblock \emph{\bibinfo{journal}{Phys. Rev. B}} \textbf{\bibinfo{volume}{6}},
  \bibinfo{pages}{1226} (\bibinfo{year}{1972}).

\bibitem{puchkov:1996}
\bibinfo{author}{Puchkov, A.~V.}, \bibinfo{author}{Basov, D.~N.} \&
  \bibinfo{author}{Timusk, T.}
\newblock \bibinfo{title}{The pseudogap state in high-\mbox{$T_c$}
  superconductors: an infrared study}.
\newblock \emph{\bibinfo{journal}{J. Phys.: Cond. Matter}}
  \textbf{\bibinfo{volume}{8}}, \bibinfo{pages}{10049} (\bibinfo{year}{1996}).

\bibitem{michon:2019}
\bibinfo{author}{Michon, B.} \emph{et~al.}
\newblock \bibinfo{title}{Thermodynamic signatures of quantum criticality in
  cuprate superconductors}.
\newblock \emph{\bibinfo{journal}{Nature}} \textbf{\bibinfo{volume}{567}},
  \bibinfo{pages}{218} (\bibinfo{year}{2019}).

\bibitem{presland:1991}
\bibinfo{author}{Presland, M.~R.}, \bibinfo{author}{Tallon, J.~L.},
  \bibinfo{author}{Buckley, R.~G.}, \bibinfo{author}{Liu, R.} \&
  \bibinfo{author}{Flower, N.}
\newblock \bibinfo{title}{General trends in oxygen stoichiometry effects on
  \mbox{$T_c$} in bi and tl superconductors}.
\newblock \emph{\bibinfo{journal}{Physica C: Superconductivity}}
  \textbf{\bibinfo{volume}{176}}, \bibinfo{pages}{95} (\bibinfo{year}{1991}).

\bibitem{hwang:2004}
\bibinfo{author}{Hwang, J.}, \bibinfo{author}{Timusk, T.} \&
  \bibinfo{author}{Gu, G.~D.}
\newblock \bibinfo{title}{High-transition-temperature superconductivity in the
  absence of the magnetic-resonance mode}.
\newblock \emph{\bibinfo{journal}{Nature (London)}}
  \textbf{\bibinfo{volume}{427}}, \bibinfo{pages}{714} (\bibinfo{year}{2004}).

\bibitem{hwang:2007a}
\bibinfo{author}{Hwang, J.}, \bibinfo{author}{Timusk, T.} \&
  \bibinfo{author}{Gu, G.~D.}
\newblock \bibinfo{title}{Doping dependent optical properties of
  \mbox{Bi$_2$Sr$_2$CaCu$_2$O$_{8+\delta}$}}.
\newblock \emph{\bibinfo{journal}{J. Phys.: Condens. Matter}}
  \textbf{\bibinfo{volume}{19}}, \bibinfo{pages}{125208}
  (\bibinfo{year}{2007}).

\bibitem{johnson:2001}
\bibinfo{author}{Johnson, P.~D.} \emph{et~al.}
\newblock \bibinfo{title}{Doping and temperature dependence of the mass
  enhancement observed in the cuprate
  \mbox{Bi$_2$⁢Sr$_2$⁢CaCu$_2$⁢O$_{8+\delta}$}}.
\newblock \emph{\bibinfo{journal}{Phys. Rev. Lett.}}
  \textbf{\bibinfo{volume}{87}}, \bibinfo{pages}{177007}
  (\bibinfo{year}{2001}).

\bibitem{carbotte:1999}
\bibinfo{author}{Carbotte, J.~P.}, \bibinfo{author}{Schachinger, E.} \&
  \bibinfo{author}{Basov, D.~N.}
\newblock \bibinfo{title}{Coupling strength of charge carriers to spin
  fluctuations in high-temperature superconductors}.
\newblock \emph{\bibinfo{journal}{Nature (London)}}
  \textbf{\bibinfo{volume}{401}}, \bibinfo{pages}{354} (\bibinfo{year}{1999}).

\bibitem{hwang:2006}
\bibinfo{author}{Hwang, J.} \emph{et~al.}
\newblock \bibinfo{title}{\mbox{a}-axis optical conductivity of detwinned
  ortho-\mbox{II} \mbox{YBa$_2$Cu$_3$O$_{6.50}$}}.
\newblock \emph{\bibinfo{journal}{Phys. Rev. B}} \textbf{\bibinfo{volume}{73}},
  \bibinfo{pages}{014508} (\bibinfo{year}{2006}).

\bibitem{hwang:2008c}
\bibinfo{author}{Hwang, J.} \emph{et~al.}
\newblock \bibinfo{title}{Bosonic spectral density of epitaxial thin-film
  \mbox{La$_{1.83}$Sr$_{0.17}$CuO$_4$} superconductors from infrared
  conductivity measurements}.
\newblock \emph{\bibinfo{journal}{Phys. Rev. Lett.}}
  \textbf{\bibinfo{volume}{100}}, \bibinfo{pages}{137005}
  (\bibinfo{year}{2008}).

\bibitem{ma:2006}
\bibinfo{author}{Ma, Y.~C.} \& \bibinfo{author}{Wang, N.~L.}
\newblock \bibinfo{title}{Infrared scattering rate of overdoped
  \mbox{Tl$_2$Ba$_2$CuO$_{6+\delta}$}}.
\newblock \emph{\bibinfo{journal}{Phys. Rev. B}} \textbf{\bibinfo{volume}{73}},
  \bibinfo{pages}{144503} (\bibinfo{year}{2006}).

\bibitem{auvray:2019}
\bibinfo{author}{Auvray, N.} \emph{et~al.}
\newblock \bibinfo{title}{Nematic fluctuations in the cuprate superconductor
  \mbox{Bi$_2$Sr$_2$CaCu$_2$O$_{8+\delta}$}}.
\newblock \emph{\bibinfo{journal}{Nat. Comm.}} \textbf{\bibinfo{volume}{10}},
  \bibinfo{pages}{5209} (\bibinfo{year}{2019}).

\bibitem{mirzaei:2013}
\bibinfo{author}{Mirzaei, S.~I.} \emph{et~al.}
\newblock \bibinfo{title}{Spectroscopic evidence for fermi liquid-like energy
  and temperature dependence of the relaxation rate in the pseudogap phase of
  the cuprates}.
\newblock \emph{\bibinfo{journal}{PNAS}} \textbf{\bibinfo{volume}{110}},
  \bibinfo{pages}{5774} (\bibinfo{year}{2013}).

\bibitem{hartnoll:2015}
\bibinfo{author}{Hartnoll, S.~A.} \& \bibinfo{author}{Karch, A.}
\newblock \bibinfo{title}{Scaling theory of the cuprate strange metals}.
\newblock \emph{\bibinfo{journal}{Phys. Rev. B}} \textbf{\bibinfo{volume}{91}},
  \bibinfo{pages}{155126} (\bibinfo{year}{2015}).

\bibitem{kamatani:2022}
\bibinfo{author}{Kamatani, T.}, \bibinfo{author}{Kitamura, S.},
  \bibinfo{author}{Tsuji, N.}, \bibinfo{author}{Shimano, R.} \&
  \bibinfo{author}{Morimoto, T.}
\newblock \bibinfo{title}{Optical response of the leggett mode in multiband
  superconductors in the linear response regime}.
\newblock \emph{\bibinfo{journal}{Phys. Rev. B}}
  \textbf{\bibinfo{volume}{105}}, \bibinfo{pages}{094520}
  (\bibinfo{year}{2022}).

\bibitem{du:2020}
\bibinfo{author}{Du, Z.} \emph{et~al.}
\newblock \bibinfo{title}{Imaging the energy gap modulations of the cuprate
  pair-density-wave state}.
\newblock \emph{\bibinfo{journal}{Nature}} \textbf{\bibinfo{volume}{580}},
  \bibinfo{pages}{65} (\bibinfo{year}{2020}).

\bibitem{wang:2021}
\bibinfo{author}{Wang, S.} \emph{et~al.}
\newblock \bibinfo{title}{Scattering interference signature of a pair density
  wave state in the cuprate pseudogap phase}.
\newblock \emph{\bibinfo{journal}{Nat. Comm.}} \textbf{\bibinfo{volume}{12}},
  \bibinfo{pages}{6087} (\bibinfo{year}{2021}).

\bibitem{nagashima:2025}
\bibinfo{author}{Nagashima, R.}, \bibinfo{author}{Mouilleron, T.} \&
  \bibinfo{author}{Tsuji, N.}
\newblock \bibinfo{title}{Optically active higgs and leggett modes in multiband
  pair-density-wave superconductors with \mbox{Lifshitz} invariant}.
\newblock \emph{\bibinfo{journal}{Phys. Rev. B}}
  \textbf{\bibinfo{volume}{112}}, \bibinfo{pages}{024503}
  (\bibinfo{year}{2025}).

\bibitem{kaminski:2006}
\bibinfo{author}{Kaminski, A.} \emph{et~al.}
\newblock \bibinfo{title}{Change of fermi-surface topology in
  \mbox{Bi$_2$Sr$_2$CaCu$_2$O$_{8+\delta}$} with doping}.
\newblock \emph{\bibinfo{journal}{Phys. Rev. B}} \textbf{\bibinfo{volume}{73}},
  \bibinfo{pages}{174511} (\bibinfo{year}{2006}).

\bibitem{loret:2018}
\bibinfo{author}{Loret, B.} \emph{et~al.}
\newblock \bibinfo{title}{Raman and arpes combined study on the connection
  between the existence of the pseudogap and the topology of the fermi surface
  in \mbox{Bi$_2$Sr$_2$CaCu$_2$O$_{8+\delta}$}}.
\newblock \emph{\bibinfo{journal}{Phys. Rev. B}} \textbf{\bibinfo{volume}{97}},
  \bibinfo{pages}{174521} (\bibinfo{year}{2018}).

\bibitem{gupta:2021}
\bibinfo{author}{Gupta, N.~K.} \emph{et~al.}
\newblock \bibinfo{title}{Vanishing nematic order beyond the pseudogap phase in
  overdoped cuprate superconductors}.
\newblock \emph{\bibinfo{journal}{PNAS}} \textbf{\bibinfo{volume}{118}},
  \bibinfo{pages}{2106881118} (\bibinfo{year}{2021}).

\bibitem{carbotte:2011}
\bibinfo{author}{Carbotte, J.~P.}, \bibinfo{author}{Timusk, T.} \&
  \bibinfo{author}{Hwang, J.}
\newblock \bibinfo{title}{Bosons in high-temperature superconductors: an
  experimental survey}.
\newblock \emph{\bibinfo{journal}{Reports on Progress in Physics}}
  \textbf{\bibinfo{volume}{74}}, \bibinfo{pages}{066501}
  (\bibinfo{year}{2011}).

\bibitem{bourges:2005}
\bibinfo{author}{Bourges, P.} \emph{et~al.}
\newblock \bibinfo{title}{The resonant magnetic mode: A common feature of
  high-\mbox{$T_c$} superconductors}.
\newblock \emph{\bibinfo{journal}{Physica C: Superconductivity}}
  \textbf{\bibinfo{volume}{424}}, \bibinfo{pages}{45} (\bibinfo{year}{2005}).

\bibitem{yu:2009}
\bibinfo{author}{Yu, G.}, \bibinfo{author}{Li, Y.}, \bibinfo{author}{Motoyama,
  E.~M.} \& \bibinfo{author}{Greven, M.}
\newblock \bibinfo{title}{A universal relationship between magnetic resonance
  and superconducting gap in unconventional superconductors}.
\newblock \emph{\bibinfo{journal}{Nat. Phys.}} \textbf{\bibinfo{volume}{5}},
  \bibinfo{pages}{873} (\bibinfo{year}{2009}).

\bibitem{cooper:2009}
\bibinfo{author}{Cooper, R.~A.} \emph{et~al.}
\newblock \bibinfo{title}{Anomalous criticality in the electrical resistivity
  of \mbox{La$_{2--x}$Sr$_x$CuO$_4$}}.
\newblock \emph{\bibinfo{journal}{Science}} \textbf{\bibinfo{volume}{323}},
  \bibinfo{pages}{603} (\bibinfo{year}{2009}).

\bibitem{homes:1993}
\bibinfo{author}{Homes, C.~C.}, \bibinfo{author}{Reedyk, M.~A.},
  \bibinfo{author}{Crandles, D.~A.} \& \bibinfo{author}{Timusk, T.}
\newblock \bibinfo{title}{Technique for measuring the reflectance of irregular,
  submillimeter-sized samples}.
\newblock \emph{\bibinfo{journal}{Appl. Opt.}} \textbf{\bibinfo{volume}{32}},
  \bibinfo{pages}{2976} (\bibinfo{year}{1993}).

\bibitem{tanner:2019}
\bibinfo{author}{Tanner, D.~B.}
\newblock \emph{\bibinfo{title}{Optical effects in solids}}
  (\bibinfo{publisher}{Cambridge Univ. Press}, \bibinfo{year}{2019}).

\bibitem{hwang:2007}
\bibinfo{author}{Hwang, J.}, \bibinfo{author}{Timusk, T.},
  \bibinfo{author}{Schachinger, E.} \& \bibinfo{author}{Carbotte, J.~P.}
\newblock \bibinfo{title}{Evolution of the bosonic spectral density of the
  high-temperature superconductor \mbox{Bi$_2$Sr$_2$CaCu$_2$O$_{8+\delta}$}}.
\newblock \emph{\bibinfo{journal}{Phys. Rev. B}} \textbf{\bibinfo{volume}{75}},
  \bibinfo{pages}{144508} (\bibinfo{year}{2007}).

\end{thebibliography}

\newpage

\setcounter{figure}{0}

\begin{figure}[!htbp]
\renewcommand{\figurename}{Extended Data Fig.}
\begin{center}
  \vspace*{-0.3 cm}%
\includegraphics[width=0.9 \columnwidth]{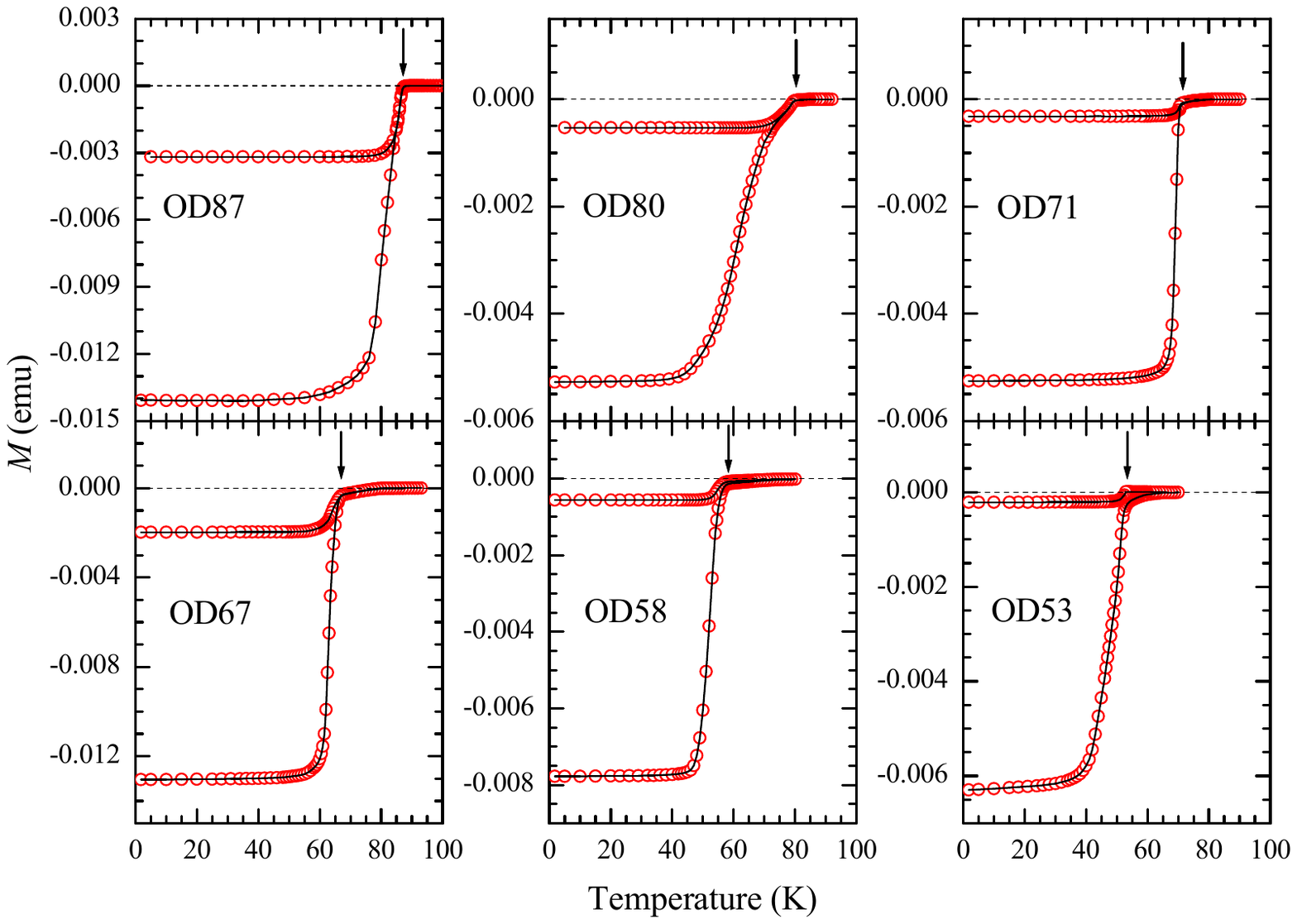}
  \vspace*{-1.0 cm}%
\caption{{\bf Measured magnetization.} The $T$-dependent measured magnetization curves near $T_c$ of all six Bi-2212 samples for the field and zero-field cooling. The vertical arrows indicate the SC transition temperatures, $T_c$. A physical property measurement system (Quantum Design PPMS, US) was utilized to obtain magnetization curves.
\label{figS1}}
\end{center}
\end{figure} 

\newpage

\begin{figure}[!htbp]
\renewcommand{\figurename}{Extended Data Fig.}
\begin{center}
  \vspace*{-0.3 cm}%
\includegraphics[width=1.0 \columnwidth]{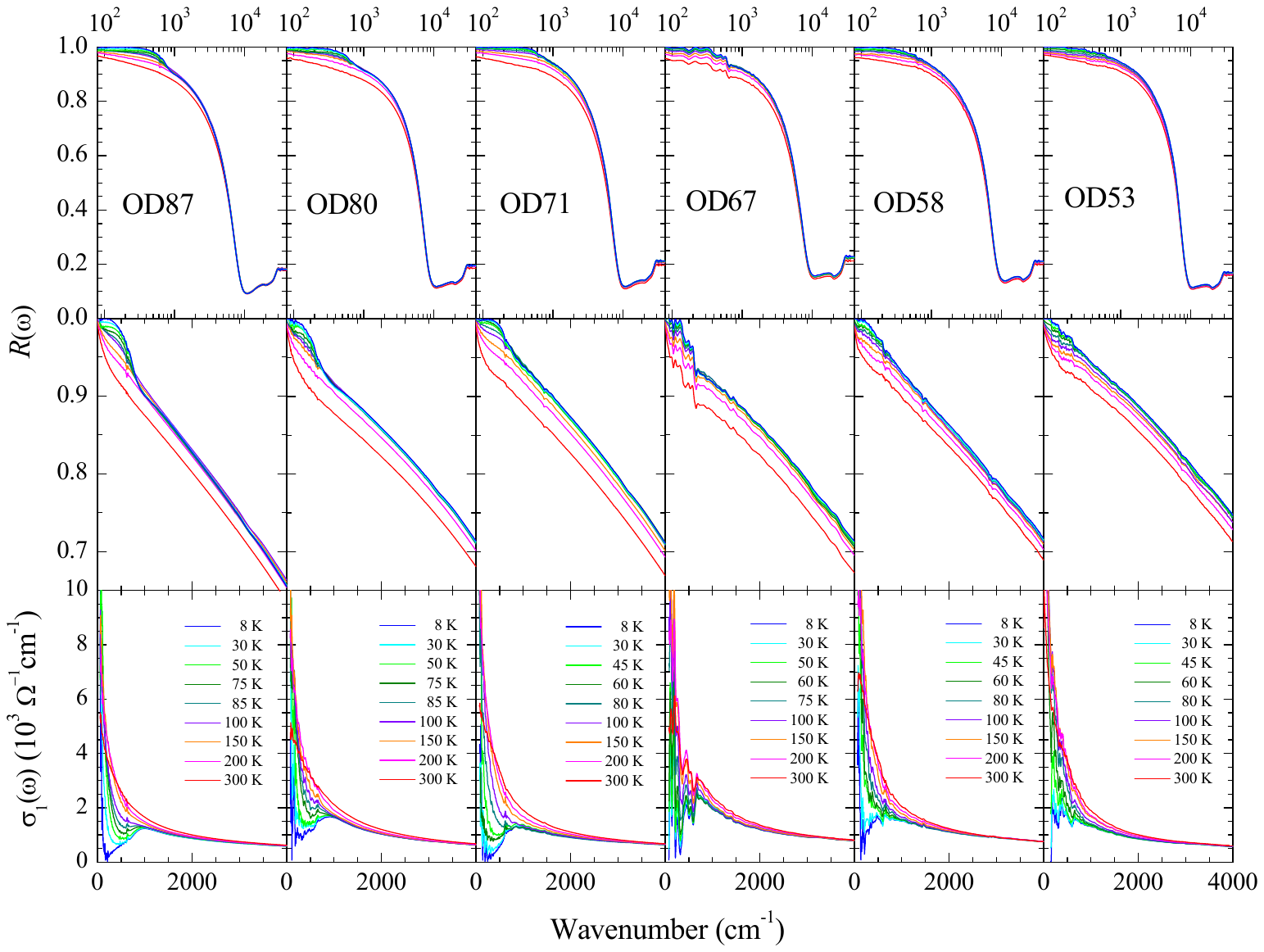}
  \vspace*{-1.5 cm}%
\caption{{\bf Measured $ab$-plane reflectance and optical conductivity.} (Top and middle panels) Measured reflectance spectra of the six Bi-2212 samples at various temperatures above and below $T_c$ in two different spectral ranges. (Bottom panels) Corresponding real parts of the optical conductivity.
\label{figS2}}
\end{center}
\end{figure} 

\newpage

\begin{figure}[!htbp]
\renewcommand{\figurename}{Extended Data Fig.}
\begin{center}
  \vspace*{-0.3 cm}%
\includegraphics[width=1.0 \columnwidth]{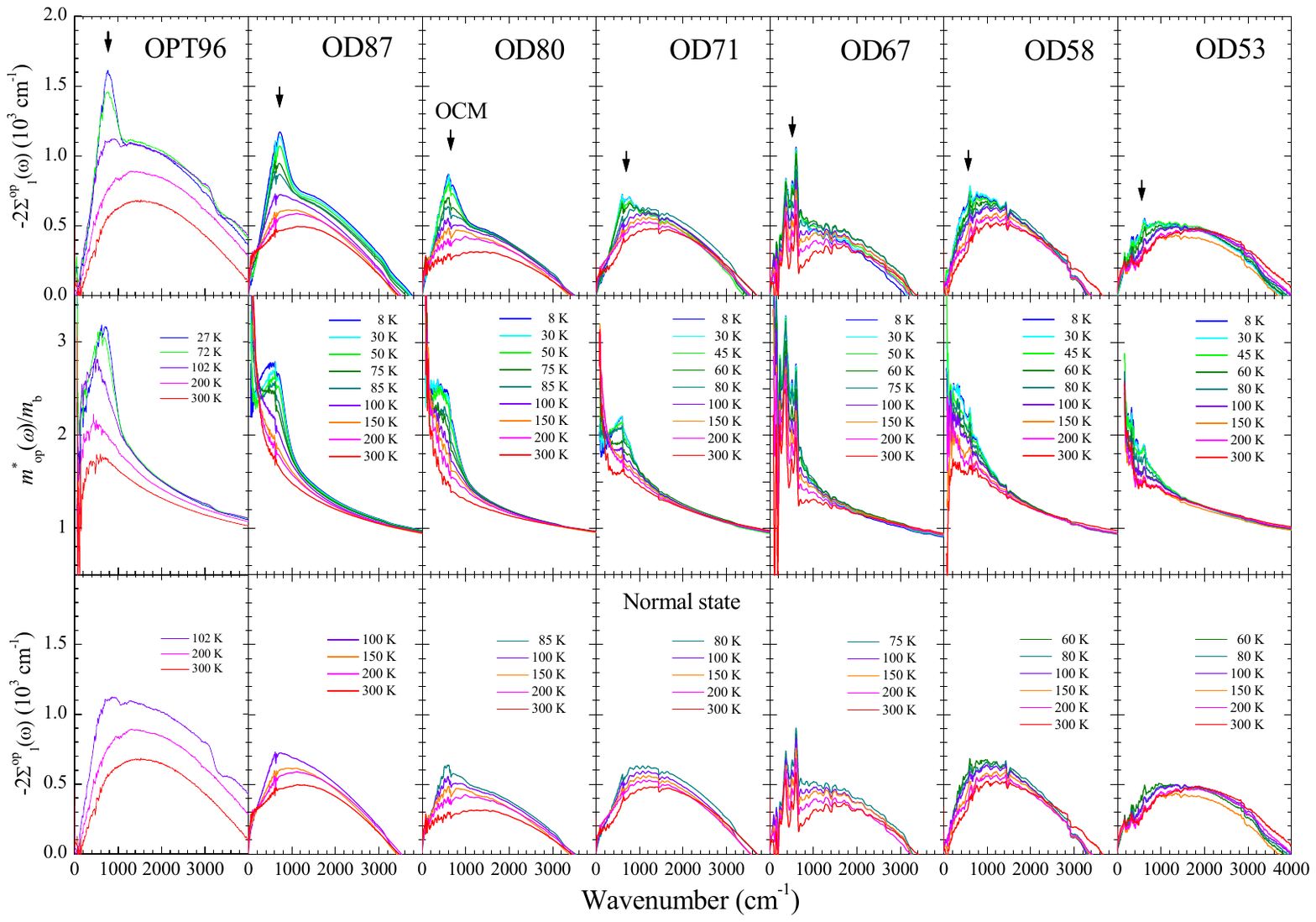}
  \vspace*{-1.5 cm}%
\caption{{\bf Real parts of optical self-energy and optical effective mass.} (Top panels) $-2\Sigma^{op}_1(\omega, T)$ of seven B-2212 samples, including an optimally doped one from a previous study \cite{hwang:2007a}, at all measured temperatures. The vertical arrows indicate the OCM. (Middle panels) Corresponding optical effective mass, $m^*_{op}(\omega)/m_b$. Note that the effective mass exhibits a corresponding feature to the OCM in a similar frequency region. (Bottom panels) $-2\Sigma^{op}_1(\omega, T)$ of the same samples only for the normal state, to show the doping-dependent evolution of $-2\Sigma^{op}_1(\omega, T)$ in the normal state separately. 
\label{figS3}}
\end{center}
\end{figure}

\newpage

\begin{figure}[!htbp]
\renewcommand{\figurename}{Extended Data Fig.}
\begin{center}
  \vspace*{-0.3 cm}%
\includegraphics[width=1.0 \columnwidth]{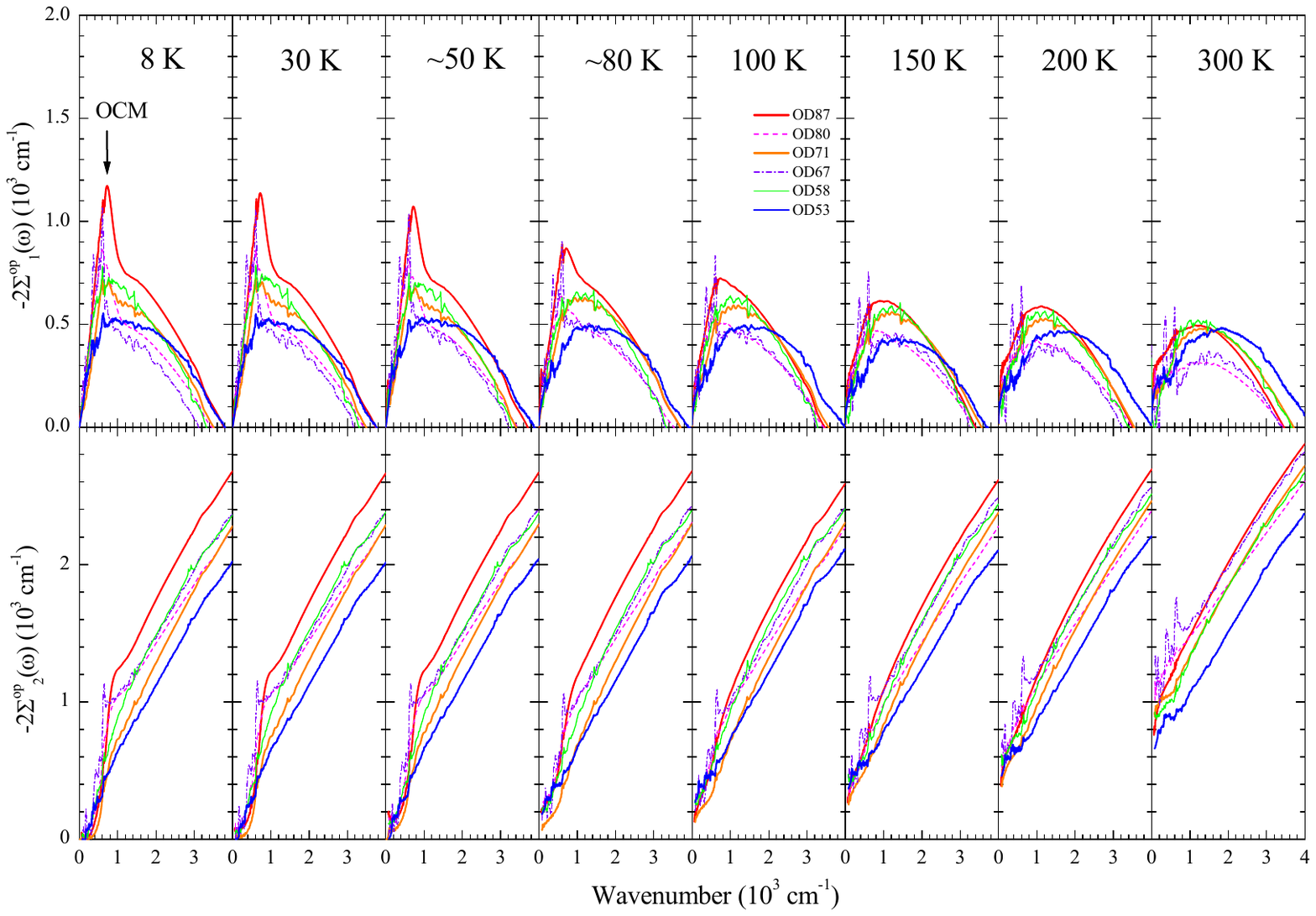}
  \vspace*{-1.5 cm}%
\caption{{\bf $T$-dependent evolution of optical self-energy} (Upper panels) Real parts of optical self-energy of six Bi-2212 samples at given temperatures. As temperature increases, the OCM decreases and eventually disappears, being consistent with the phase diagram in Fig. \ref{fig3}b. (Lower panels) Corresponding imaginary parts of optical self-energy of six Bi-2212 samples. Note that the overall level of the optical scattering rate decreases as doping increases, which might be associated with the doping-induced reduction of spin fluctuations.
\label{figS4}}
\end{center}
\end{figure}

\newpage

\begin{figure}[!htbp]
\renewcommand{\figurename}{Extended Data Fig.}
\begin{center}
  \vspace*{-0.3 cm}%
\includegraphics[width=0.7 \columnwidth]{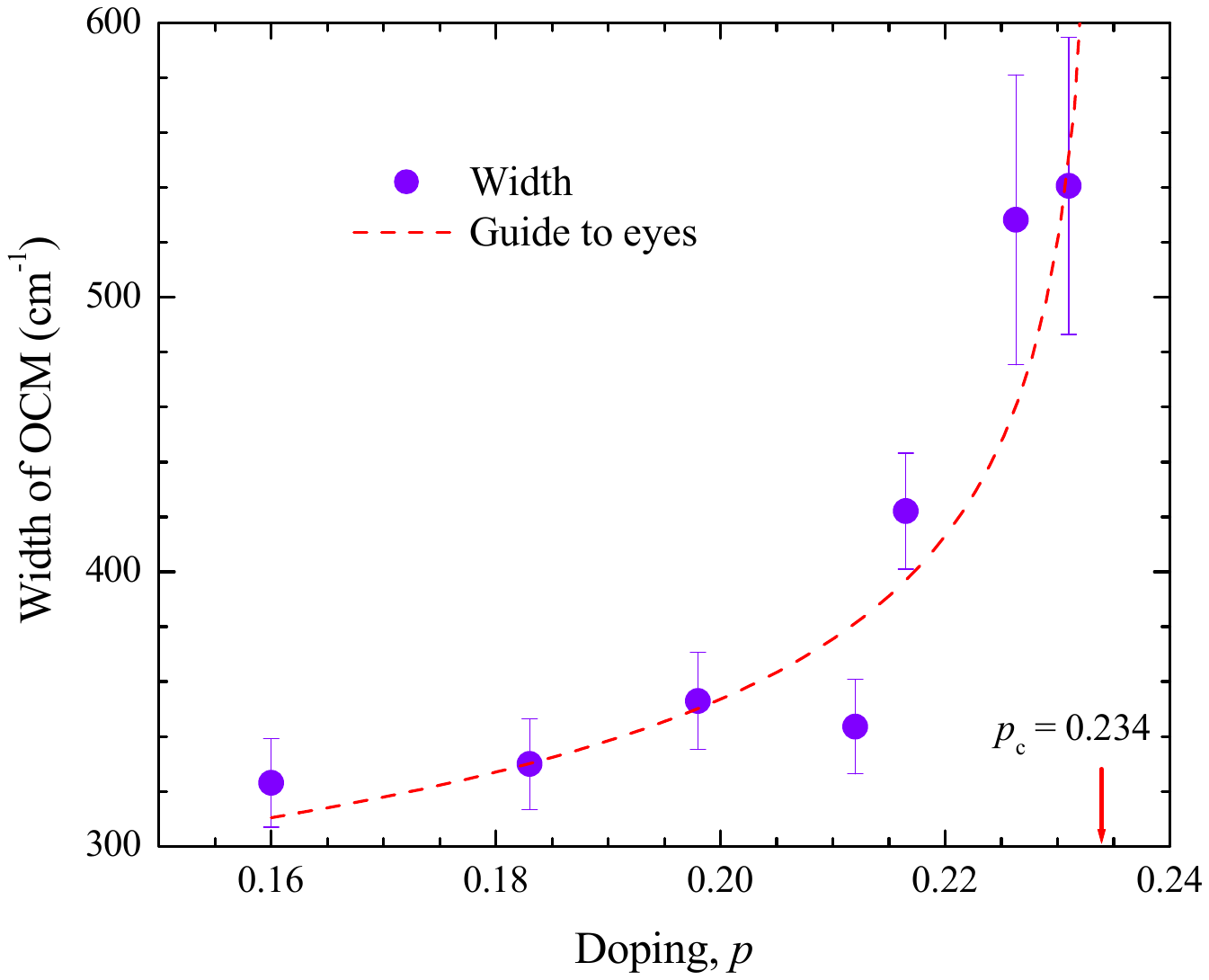}
  \vspace*{-0.8 cm}%
\caption{{\bf Doping-dependent width of OCM.} The width of the OCM rapidly widens as the doping approaches the critical doping ($p_c \approx$ 0.234).
\label{figS5}}
\end{center}
\end{figure}

\newpage

\begin{figure}[!htbp]
\renewcommand{\figurename}{Extended Data Fig.}
\begin{center}
  \vspace*{-0.3 cm}%
\includegraphics[width=0.8 \columnwidth]{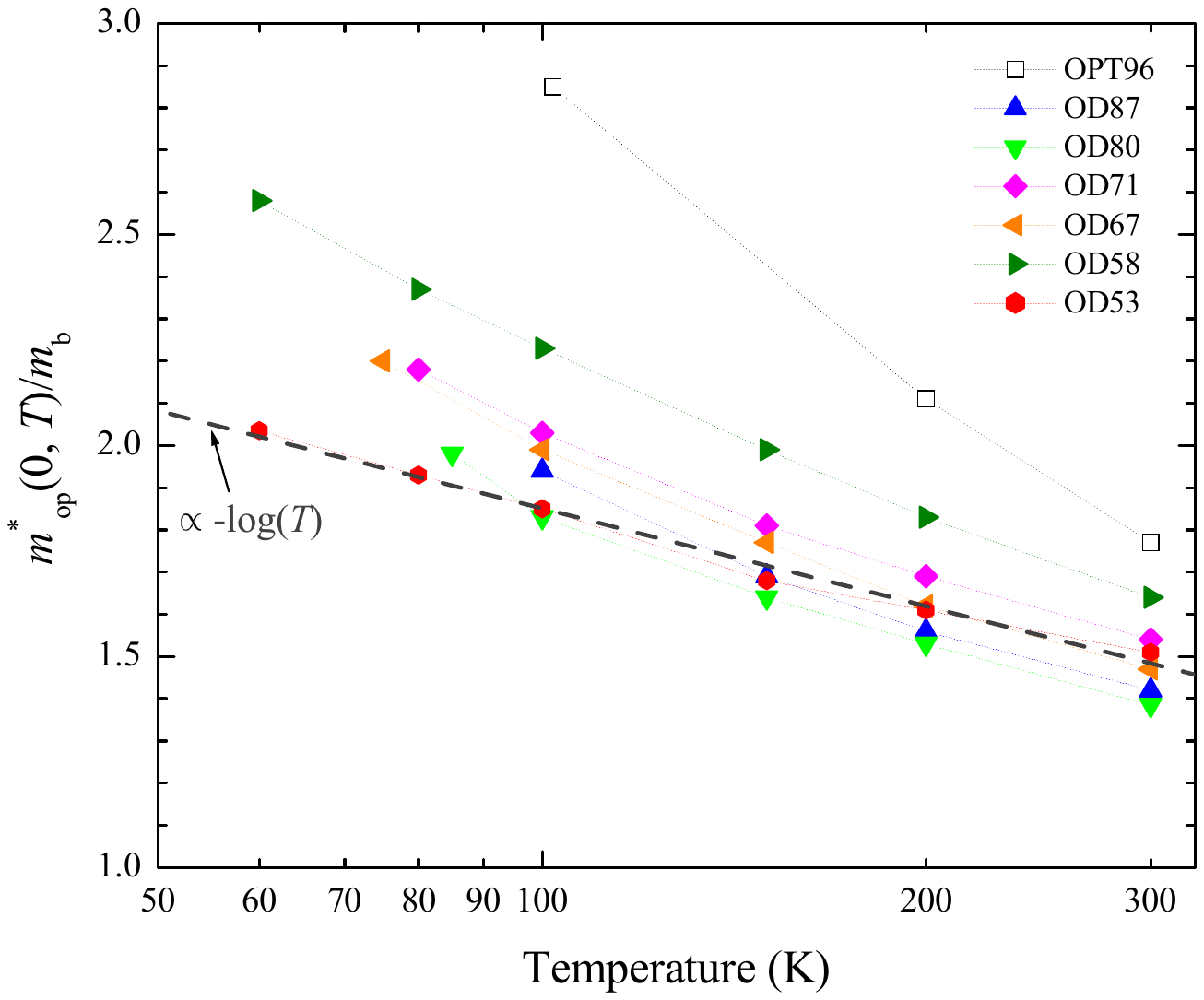}
  \vspace*{-1.0 cm}%
\caption{{\bf Effective mass at zero frequency.} $m^*_{op}(0, T)/m_b$ of all seven Bi-2212 samples, including the optimally doped one from a previous study \cite{hwang:2007a}, in semi-log scale.}
\label{figS6}
\end{center}
\end{figure}

\newpage

\begin{figure}[!htbp]
\renewcommand{\figurename}{Extended Data Fig.}
\begin{center}
  \vspace*{-0.3 cm}%
\includegraphics[width=0.8 \columnwidth]{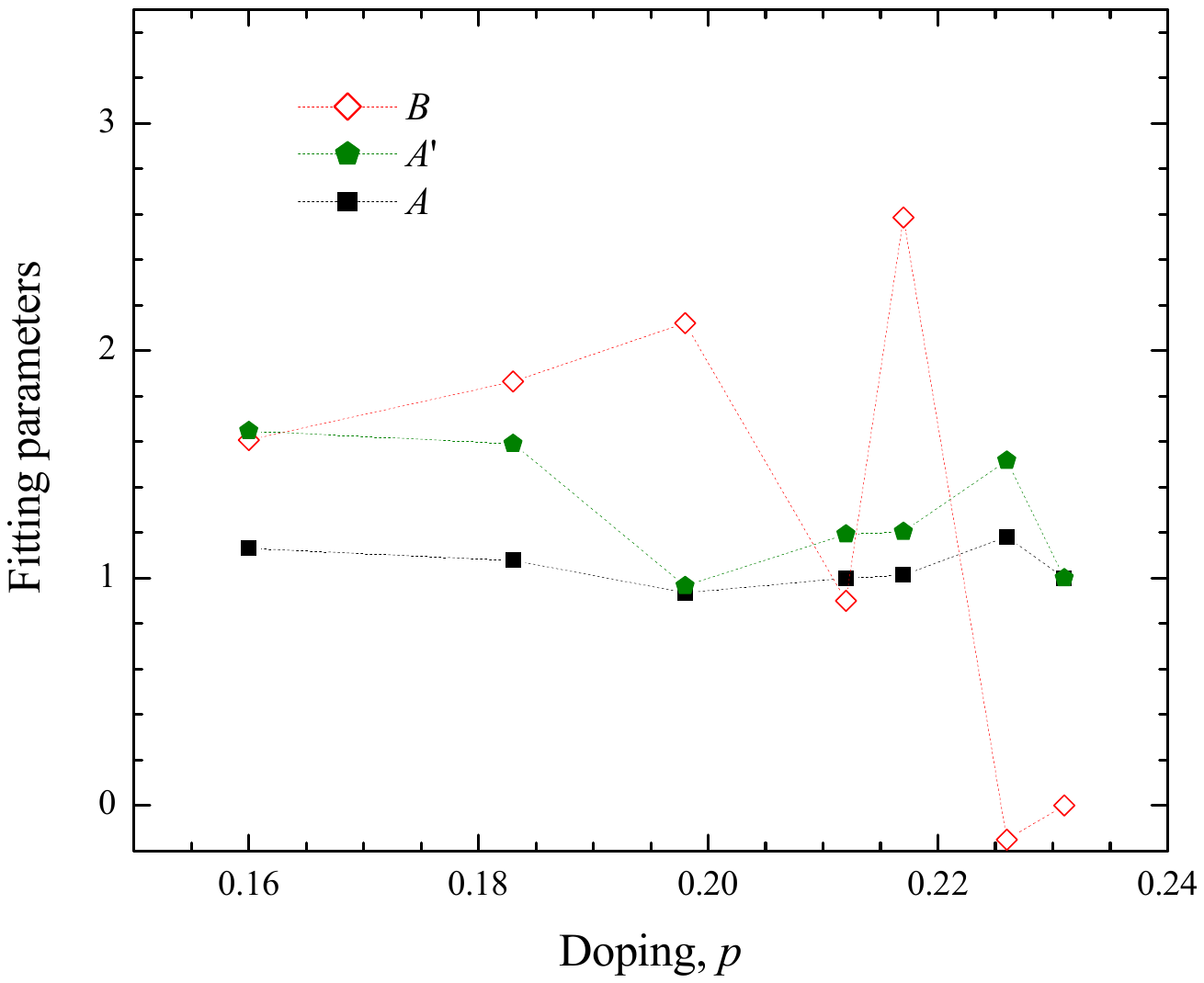}
  \vspace*{-1.0 cm}%
\caption{{\bf Fitting parameters of the optical scattering rate and effective mass with scaling functions.} The fitting parameters ($A$, $B$, and $A'$) of the optical scattering rate and effective mass with the scaling functions, $f_{\tau}(S)$ and $f_m(S)$ (see Fig. \ref{fig3}a). Note that $f_{\tau}(S)$ and $f_m(S)$ are the scaling functions at $p$ = 0.231. $A$ is close to 1.0 at all doping and $A'$ exhibits more doping dependence and is approximately 1.4 times larger than $A$.
\label{figS7}}
\end{center}
\end{figure}

\end{document}